\documentclass[12pt,preprint]{aastex}

\shorttitle{Thick Disk Model for ULSs}
\shortauthors{Gu et al.}

\begin{document}

\title{Thick Accretion Disk Model for Ultraluminous Supersoft Sources}

\author{Wei-Min Gu\altaffilmark{1,4}, Mou-Yuan Sun\altaffilmark{1},
You-Jun Lu\altaffilmark{2,5}, Feng Yuan\altaffilmark{3,1,4},
and Ji-Feng Liu\altaffilmark{2,5}}

\altaffiltext{1}{Department of Astronomy and Institute of Theoretical Physics
and Astrophysics, Xiamen University, Xiamen, Fujian 361005, China;
guwm@xmu.edu.cn}
\altaffiltext{2}{Key Laboratory of Optical Astronomy, National Astronomical
Observatories, Chinese Academy of Sciences, 20A Datun Road, Chaoyang District,
Beijing 100012, China}
\altaffiltext{3}{Shanghai Astronomical Observatory, Chinese Academy
of Sciences, 80 Nandan Road, Shanghai 200030, China}
\altaffiltext{4}{SHAO-XMU Joint Center for Astrophysics, Xiamen University,
Xiamen, Fujian 361005, China}
\altaffiltext{5}{College of Astronomy and Space Sciences, University of
Chinese Academy of Sciences, 19A Yuquan Road, Beijing 100049, China}

\begin{abstract}
We propose a geometrically thick, super-Eddington accretion disk model,
where an optically thick wind is not necessary,
to understand ultraluminous supersoft sources (ULSs).
For high mass accretion rates $\dot M \ga 30\dot M_{\rm Edd}$
and not small inclination angles $\theta \ga 25^{\circ}$,
where $\dot M_{\rm Edd}$ is the Eddington accretion rate,
the hard photons from the hot inner region may be shaded by the
geometrically thick inner disk, and therefore only the soft photons
from the outer thin disk and the outer photosphere of the thick disk
can reach the observer.
Our model can naturally explain the approximate relation between
the typical thermal radius and the thermal temperature,
$R_{\rm bb} \propto T_{\rm bb}^{-2}$.
Moreover, the thick disk model can unify ULSs and normal
ultraluminous X-ray sources, where the different observational
characteristics are probably related to
the inclination angle and the mass accretion rate.
By comparing our model with the optically thick outflow model,
we find that less mass accretion rate is required in our model.
\end{abstract}

\keywords{accretion, accretion disks --- black hole physics
--- X-rays: binaries}

\section{Introduction}

Ultraluminous supersoft sources (ULSs) are a particular group of X-ray binaries,
with both a high luminosity of $\sim$ a few $10^{39}$~erg~s$^{-1}$ and
a supersoft thermal spectrum with a peak temperature $\la 0.1$~keV
\citep{Stefano03}.
Recently, significant progress has been achieved for two individual ULSs,
M101 X-1 and M81 ULS-1. For M101 X-1, \citet{Liu13} 
confirmed that the system contains a Wolf-Rayet star and revealed that
the orbital period is 8.2 days. Based on this orbital period, they
proposed that the black hole probably has a mass around
$20-30 M_{\sun}$. This is the first time to measure the black hole mass
in a ULS by dynamic method. For another ULS, M81 ULS-1, \citet{Liu15}
found blueshifted, broad H$\alpha$ emission lines,
which indicate that there exists a relativistic baryonic jet.
Moreover, the blueshifted value reveals the inclination angle
$\theta < 60^{\circ}$. Consequently,
since a white dwarf system is unable to launch such a relativistic jet,
the model based on a white dwarf can be ruled out.

Two well-known models for ULSs are related to intermediate-mass
black holes (IMBHs) and stellar-mass black holes, respectively.
The model based on a standard thin disk \citep{SS73} around an IMBH can
explain both high bolometric luminosity $L_{\rm bol}$
and low inner disk temperature $T_{\rm in}$ due to the relation
$T_{\rm in} \propto M_{\rm BH}^{-1/4}$. However, compared with
the well-known behaviors of Galactic black hole X-ray binaries (BHXBs),
such a model may confront two difficulties.
The first one is related to the relative luminosity.
Usually, the disk-dominated thermal state of a BHXB occurs
for $0.02L_{\rm Edd} \la L_{\rm bol} \la 0.3L_{\rm Edd}$
\citep{Maccarone03,McClintock06}, where $L_{\rm Edd}$ is
the Eddington luminosity. Such a range corresponds to
$0.02 \la \dot m \la 0.3$, where $\dot m \equiv \dot M/\dot M_{\rm Edd}$ 
is the dimensionless mass accretion rate, and $\dot M_{\rm Edd}$ is the
Eddington accretion rate defined as $L_{\rm Edd}/(\eta c^2)$.
In the present work we choose $\eta = 1/16$ corresponding to
the well-known Paczy{\'n}sky-Wiita potential \citep{PW80}.
For the IMBH model, however, the relative luminosity $L_{\rm bol}/L_{\rm Edd}$
is generally far below the lower limit 0.02.
The second difficulty is related to the relativistic jet.
The existence of jet in M81 ULS-1 indicates that the jet may be a common
phenomenon in ULSs. In Galactic BHXBs, there exist two
types of jets, i.e., the steady jet in the low/hard state, and the episodic
jet during the transition between low/hard state and high/soft state
\citep{Fender04}.
However, a standard thin disk around an IMBH should correspond to a purely
soft state, which disagrees with the conditions for the above two types of jets.
Thus, the IMBH model can be ruled out.

The other model, which is based on a stellar-mass black hole,
may also confront difficulties. First, for such a black hole, the accretion rate
ought to be around or super Eddington due to the high luminosity.
In our Galaxy, however, such a supersoft spectrum has never been found among
the around twenty identified BHXBs, some of which are likely to achieve or above
Eddington luminosity such as GRS~1915+105. Second, many ultraluminous
X-ray sources (ULXs)
have been found in nearby galaxies, most of which are believed to be
stellar-mass black holes with super-Eddington accretion
\citep[for a review, see][]{Feng11}. If ULSs are
in the same scenario, then a question will arise: how can the super-Eddington
accretion around stellar-mass black hole show the apparently different
radiation characteristics?

In this $Letter$, we will propose a new model to understand ULSs, and try
to unify ULSs and ULXs in the same frame. The remainder of this $Letter$
is organized as follows. The thick accretion disk model is proposed
in Section~2. Application of such a model to ULSs is presented in Section~3.
Conclusions and discussion are made in Section~4.

\section{Thick disk model}

As mentioned in the previous section, ULSs are likely to be powered by
super-Eddington accretion around stellar-mass black holes.
The classic model for super-Eddington accretion is the slim disk
\citep{Abram88}. The half-thickness $H$ of slim disks approaches the radius
$R$, which are geometrically much thicker than the standard thin disks.
Recent simulations have made great progress on the
super-Eddington accretion process, including the identification of
an important new energy transport mechanism in addition to the diffusion,
i.e., the vertical advection of radiation \citep{Jiang14,Sadow15b},
the radiation-powered baryonic jet \citep{Sadow15a},
the strongly anisotropic feature of radiation \citep{Sadow15b,Narayan15},
and the presence of strong wind
\citep{Ohsuga11,Yang14,Sadow15a,Sadow15b,Moller15}.
In addition, such simulations showed that the disk is geometrically
thick, i.e., the opening angle between the photosphere and the
equatorial plane is quite large.
For example, \citet{Narayan15} made global simulations for the case
$\dot m =11$ and showed a geometrically thick accretion disk, where the
polar angle of the photosphere is around $25^{\circ}$.
As a consequence, the inner disk will be invisible to
an observer with an inclination angle $\theta \ga 25^{\circ}$.
On the other hand, some analytic works also pointed out that the super-Eddington
accretion disk, which is optically thick and probably advection dominated,
is likely to be geometrically thick rather than slim if the gravitational force
of the central black hole is well treated \citep{Gu07,Gu09,Gu12}.

In the present work, we propose a geometrically thick,
super-Eddington accretion model for ULSs, where the hot inner disk can
be shaded and the hard photons from this part cannot be observed
if the inclination angle is not small.
The high bolometric luminosity and the low thermal temperature of ULSs
suggest a large photosphere radius,
$R_{\rm bb} \ga 100 R_{\rm g}$, where $R_{\rm g} \equiv 2GM_{\rm BH}/c^2$
is the Swarzschild radius.
Thus, the mass accretion rate ought to be extremely high such that
the flow can have a geometrically thick inner disk extended to
$\ga 100 R_{\rm g}$.
In this scenario, there are two necessary conditions for ULSs, i.e.,
high mass accretion rates and not small inclination angles.
We would stress that a basic assumption in our model is that
an optically thick wind does not exist, which is unnecessary to
explain the observational properties of ULSs.
A cartoon picture of such a model, which includes ULSs and ULXs,
is shown in Figure~1. It is seen from this figure that the accretion disk
can be separated into two parts by a typical transition radius $R_{\rm tr}$,
i.e., an inner thick disk ($R \la R_{\rm tr}$) and an outer thin disk
($R \ga R_{\rm tr}$).
In the classic theory for super-Eddington accretion, there exists a
transition radius which connects an inner slim disk and an outer
thin disk \citep[e.g.,][]{Abram88,Watarai00}.
As mentioned in the previous paragraph, for the super-Eddington accretion case,
the inner disk is more likely to be geometrically thick rather than slim.
That is why we propose a thick inner disk here.
In addition, the system may have a baryonic jet as revealed by \citet{Liu15}.
Such a baryonic jet was also found in simulations with a relativistic speed
$\sim 0.3c$ \citep{Sadow15a}.
Since the jet can extend to a large distance, it may be observed
even for large inclination angles. For the individual source
M81 ULS-1, as mentioned in the first section, the blueshifted
H$\alpha$ emission lines indicate that the inclination angle
ought to be $\theta < 60^{\circ}$.

On the contrary, for a small inclination angle, an observer
is able to see the hard photons from the inner disk. Moreover,
according to the simulation results \citep[e.g., Figure~17 of][]{Narayan15},
the isotropic equivalent luminosity can be far beyond the Eddington one
($\sim 10L_{\rm Edd}$ for $\theta < 25^{\circ}$,
as shown by the filled red circles in their Figure~17).
Thus, a super-Eddington accreting stellar-mass black hole may appear as a normal
ULX to the observer with small inclination angles, as illustrated by Figure~1.

It is known that the spin of stellar-mass black hole can be measured
through the X-ray continuum-fitting method \citep{Zhang97}, which is
based on the standard thin disk theory.
However, \citet{McClintock06} found that the spin parameter $a_{*}$ of
GRS~1915+105 decreases for high luminosity $L \ga 0.3L_{\rm Edd}$.
Obviously, the value of $a_{*}$ cannot have significant change in a relatively
short timescale. The reason is that the inner radius $R_{\rm in}$
moves outward for $L \ga 0.3L_{\rm Edd}$, and therefore it is larger than
the radius of the innermost stable circular orbit $R_{\rm ISCO}$.
A possible physical explanation is that the disk will inflate
for $L \ga 0.3L_{\rm Edd}$
and therefore the innermost region may be shaded \citep{McClintock06}.
We would stress that such outward moving $R_{\rm in}$ above $0.3L_{\rm Edd}$ is
not a unique phenomenon for GRS~1915+105,
which is verified as a near extreme Kerr black hole.
For another BHXB LMC~X-3 with low spin \citep{Steiner14}, it was also found
that $R_{\rm in}$ moves outward for $L \ga 0.3L_{\rm Edd}$ \citep{Steiner10}.
Even with a new spectral model ``slimbb", which is based on theoretical
works on slim disks \citep[e.g.,][]{Sadow11} and the energy advection
is taken into account, $R_{\rm in}$ still moves outward
beyond $0.3L_{\rm Edd}$ \citep{Straub11}.
In our opinion, the theoretical work \citep{Sadow11} has achieved great
progress on the slim disk model, but may not be so accurate
due to the approximate vertical component of the gravitational force.
As discussed in \citet{Gu12}, if the gravitational force is well treated
(such as using the spherical coordinates), the inner disk will be much thicker,
which may provide a clue to solve the inconsistent $a_{*}$ problem.
Moreover, the outward moving $R_{\rm in}$ was also found in neutron star
X-ray binaries \citep[e.g., XTE J1701-462,][]{Weng11}
and ULXs \citep[e.g., NGC 1313 X-2,][]{Weng14}.
Thus, we can regard $0.3L_{\rm Edd}$ as a general upper limit luminosity for
the standard thin disk, beyond which such a model may be invalid. 

Previous theoretical works \citep[e.g.,][]{Watarai00,Kato08} showed that
the typical transition radius $R_{\rm tr}$ is proportional to $\dot m$,
we therefore assume the following equation:
\begin{equation}
R_{\rm tr} = \lambda \dot m R_{\rm g} \ ,
\end{equation}
where $\lambda$ is a dimensionless parameter. The value of $\lambda$ can be
estimated as follows. For a non-spinning black hole,
we can regard $R_{\rm ISCO} = 3R_{\rm g}$ 
as the inner radius for luminosity below the critical value
$0.3L_{\rm Edd}$, which corresponds to $\dot m = 0.3$. Beyond the critical
value the inner radius will start to move outward.
Thus, $\lambda \sim 10$ may
be a good choice since it can match the above equation
($R_{\rm tr} = R_{\rm ISCO}$) at the critical point $\dot m = 0.3$.
We therefore adopt $\lambda = 10$ for the following analyses.

For a large inclination angle, since the inner disk is invisible,
the luminosity from the outer thin disk ($R \ga R_{\rm tr}$)
can be derived as
\begin{equation}
L_{\rm disk}^{\rm thin} \approx \int_{R_{\rm tr}}^{\infty}
\frac{3GM_{\rm BH}\dot M}{8\pi R^3} \cdot 4\pi R~dR = 1.2 L_{\rm Edd} \ .
\end{equation}
It is interesting that the above equation implies that $L_{\rm disk}^{\rm thin}$
is independent of $\dot m$.

As shown by Figure~1, apart from the outer thin disk, the radiation from
the outer photosphere of the thick disk also has contribution to the total
luminosity for a large inclination angle.
Since the radiative force should be less than gravitational force
at the photosphere, we may roughly assume that the former balances half of
the latter, i.e., $L_{\rm disk}^{\rm thick} \sim 0.5 L_{\rm Edd}$.
Both $L_{\rm disk}^{\rm thin}$ and $L_{\rm disk}^{\rm thick}$ are the thermal
radiation, so we may simply calculate the total bolometric luminosity as
\begin{equation}
L_{\rm bol} \approx L_{\rm disk}^{\rm thin} + L_{\rm disk}^{\rm thick}
= 1.7 L_{\rm Edd} \ ,
\end{equation}
which is also independent of $\dot m$.
We would stress that, even for the extreme value of $L_{\rm disk}^{\rm thick}$
such as 0 or $L_{\rm Edd}$, the variation of the total bolometric luminosity
$L_{\rm bol}$ is less than 30\%, which will not have
essential influence on the present results.

The typical blackbody radius $R_{\rm bb}$ and temperature $T_{\rm bb}$ should
match the relation:
\begin{equation}
L_{\rm bol} = 4\pi R_{\rm bb}^2 \sigma T_{\rm bb}^4 \ .
\end{equation}
Equation~(3) implies a saturation of $L_{\rm bol}$ for a certain
ULS. We therefore directly obtain the relation
$R_{\rm bb} \propto T_{\rm bb}^{-2}$ by Equation~(4) for varying accretion rates.
Such a relation is in good agreement with the observational data,
as revealed by \citet{Soria15b}.
More interestingly, Equation~(3) indicates
$L_{\rm bol} \propto M_{\rm BH}$, which suggests that the
central black hole mass can simply be estimated by the luminosity.
For the particular ULS M101 X-1, which is the unique source with dynamic
measurement, the black hole mass can be estimated by the luminosity
$L_{\rm bol} \approx 5\times 10^{39}$~erg~s$^{-1}$ and thus
$M_{\rm BH} \approx 23 M_{\sun}$, which agrees with the dynamic result,
i.e., probably 20-30$M_{\sun}$ \citep{Liu13}.

A unified description of ULSs, normal ULXs, and BHXBs is shown in Figure~2,
which is based on stellar-mass black hole systems. Galactic BHXBs
are generally below the Eddington luminosity. According to our model,
there are two necessary conditions for ULSs to appear, which are
high accretion rates $\dot m \ga 30$ and not small inclination angles
$\theta \ga 25^{\circ}$. For the other cases with super-Eddington
accretion, i.e., either small inclination angles $\theta \la 25^{\circ}$
or moderate super-Eddington accretion $1 \la \dot m \la 30$,
since the hot inner disk is visible to the observer and the isotropic
equivalent luminosity is beyond the Eddington one,
the sources are likely to appear as normal ULXs.
We would stress that, even though ULSs have a wider range for the
inclination angle ($\theta \ga 25^{\circ}$) than ULXs
($\theta \la 25^{\circ}$), the required mass accretion rate is much more
critical for ULSs ($\dot m \ga 30$) than for ULXs ($\dot m \ga 1$).
Thus, it is reasonable that we have found hundreds of ULXs whereas only
several ULSs.

\section{Application to ULSs}

The transition radius $R_{\rm tr}$ may be roughly regarded as the typical
blackbody radius, i.e., $R_{\rm tr} \approx R_{\rm bb}$.
This is the location where the photosphere of the inner quasi-spherical
part touches the equatorial plane.
By combining Equations~(1), (3) and (4) we can derive
the expression of $k T_{\rm bb}$ as
\begin{equation}
k T_{\rm bb} \approx 660~\dot m^{-\frac{1}{2}}
\left( \frac{M_{\rm BH}}{10M_{\sun}} \right)^{-\frac{1}{4}} \ {\rm eV} \ .
\end{equation}
Our analytic results (solid and dashed lines) together with
the observations (symbols) are shown in Figure~3.
The two horizontal blue dashed lines correspond to fixed black hole masses,
$M_{\rm BH} = 3 M_{\sun}$ and $30 M_{\sun}$,
which are based on Equation~(3).
The two inclined red solid lines correspond to fixed accretion rates
$\dot m = 30$ and $100$,
which are derived by combining Equations~(3) and (5).
The symbols represent observational results of a set of seven ULSs,
which are taken from Table~2 of \citet{Soria15b}.
It is seen from Figure~3 that most of the symbols locate
between the two blue dashed lines and also between the two red solid lines,
i.e., in the range
$3M_{\sun} \la M_{\rm BH} \la 30M_{\sun}$ and $30 \la \dot m \la 100$,
which indicates a stellar-mass black hole with extremely high accretion rates.

As mentioned in the first section, the radiation characteristic of ULSs
challenges the classic accretion theory.
Both simulations \citep[e.g.,][]{Ohsuga11,Sadow15a} and analytic studies
\citep[e.g.,][]{Gu12,Gu15}
showed that outflows are significant in super-Eddington
accretion flows. Another possible mechanism for ULSs is
the optically thick outflow model, which was first introduced by
\citet{King03} and was developed and applied to M101 X-1 by \citet{Shen15}.
Recently, \citet{Soria15a} and \citet{Soria15b} investigated such an issue
in more details based on the observational data from $Chandra$ and
$XMM$-$Newton$. It is interesting that, despite the completely
different models, the analytic results are similar between the outflow
model and ours. For instance, Equations~(24-25) of \citet{Soria15a}
also imply that the total luminosity is roughly independent of
$\dot m$, and is proportional to the black hole mass. As a consequence,
their blue solid lines in the right two panels of Figure~9 \citep{Soria15a},
which correspond to fixed black hole masses, are nearly horizontal
as shown in our Figure~3 (blue dashed lines).

For more detailed comparison, we find that the
required mass accretion rate in our model is significantly less than
that in the outflow model. For example, for the source M101 X-1,
their Figure~9 \citep{Soria15a} indicates $\dot m \ga 1000$, whereas
in our model the accretion rate is probably in the range
$30 \la \dot m \la 100$, as shown in Figure~3.
In addition, we would stress that the essential difference
from the outflow model is that an optically thick wind is not necessary
in our model to explain the observational properties of ULSs.
For the case that a strong optically thick wind exists, the transition
radius may be located much further away, i.e., $R_{\rm tr}^{'} \gg R_{\rm tr}$,
whereas the effect on the total bolometric luminosity may be quite weak
($L_{\rm bol}$ may stay close to $L_{\rm Edd}$).
As a consequence, for a given pair of the observational $L_{\rm bol}$
and $T_{\rm bb}$ ($R_{\rm bb}$ can therefore be derived by Equation~(4)),
the required $\dot m$ in our model may be even lower due to the relations
$R_{\rm tr} \propto \dot m$ (Equation (1)) and
probably $R_{\rm bb} \gg R_{\rm tr}$.

\section{Conclusions and discussion}

In this $Letter$, we have proposed a geometrically thick
accretion disk model to understand ULSs.
In our opinion, such a model can unify ULSs and most ULXs under the scenario
of a stellar-mass black hole with super-Eddington accretion.
There are two necessary conditions for ULSs, i.e.,
high mass accretion rates $\dot m \ga 30$ and not small inclination angles
$\theta \ga 25^{\circ}$. For the other cases with super-Eddington
accretion, the sources are likely to appear as normal ULXs.
Our model can naturally explain the observational relation
$R_{\rm bb} \propto T_{\rm bb}^{-2}$ for ULSs.
In addition, we suggest that the black hole mass in ULSs can be estimated
simply from the total bolometric luminosity.
By comparing our model with the optically thick outflow model,
we find that less mass accretion rate is required in our model.

We would point out that there exists uncertainty for the value of $\lambda$
in Equation~(1). In the present work we adopt $\lambda = 10$ according to
previous works on black hole spin measurement. From the theoretical point
of view, \citet{Watarai00} implies $\lambda \sim 2.4$ due to
their result $R_{\rm tr} \approx R_{\rm in}$ for $\dot m = 1.25$ (note that
there exists different definition of $\dot m$, thus their $\dot m =20$
is equivalent to our $\dot m =1.25$). Since their analyses are based on a Taylor
Expansion of the vertical component of the gravitational force, which is
probably magnified particularly for geometrically not thin disks
\citep{Gu07,Gu12}, the geometrically thickness may be underestimated and therefore
the real value of $R_{\rm tr}$ may be larger. Actually, \citet{Gu12}
investigated this issue in spherical coordinates in order to avoid the
approximation of gravity, and found that the disk can be geometrically thick
at $10R_{\rm g}$ for $\dot m = 0.6$, which corresponds to $\lambda \approx 17$
from Equation~(1).
Considering the Newtonian potential was adopted in \citet{Gu12} and therefore
the viscous heating rate has been enlarged, $\lambda$ should be less than 17.
With regard to the theoretical range $2.4 < \lambda < 17$,
it may be a reasonable assumption for $\lambda = 10$.
Nevertheless, variation of $\lambda$ by a small factor
will have only slight quantitative influence on the present results.

Our model is based on the stellar-mass black hole system. For ULXs,
however, we should mention the other two possibilities.
One possibility is neutron star X-ray binaries such as M82 X-2
\citep{Bachetti14}. The other one may be IMBH systems,
particularly for those sources with extremely high luminosity
$L \ga 10^{41}$~erg~s$^{-1}$, such as ESO 243-49 HLX-1 \citep{Farrell09}.

As discussed in Section~2, the relation $L_{\rm bol} \approx 1.7 L_{\rm Edd}$
may be used to estimate the black hole mass in ULSs.
Although the inclination angle may have influence on the apparent luminosity,
the variation should not be essential due to the limit range of the angle.
Moreover, such a relation indicates a saturated bolometric luminosity
for ULSs, or even more general, for super-Eddington accretion in different
scale. In this scenario, for super-massive black holes in active galactic nuclei,
if mass supply is sufficient, a similar supersoft state may be found.
If this is the case, such system may be regarded as another type
of ``Standard Candle" and will have potential application to
the study of cosmology.

\acknowledgments

The authors thank Shan-Shan Weng for beneficial discussions,
and the referee for constructive comments that improved the $Letter$.
This work was supported by the National Basic Research Program of China
(973 Program) under grants 2014CB845800 and 2014CB845705,
the National Natural Science Foundation of China under grants 11573023,
11333004, 11373031, 11133005, 11573051, 11425313 and 11222328,
the CAS/SAFEA International Partnership Program for Creative Research Teams,
and the Fundamental Research Funds for the Central Universities
under grant 20720140532.

\clearpage

\begin{figure}
\plotone{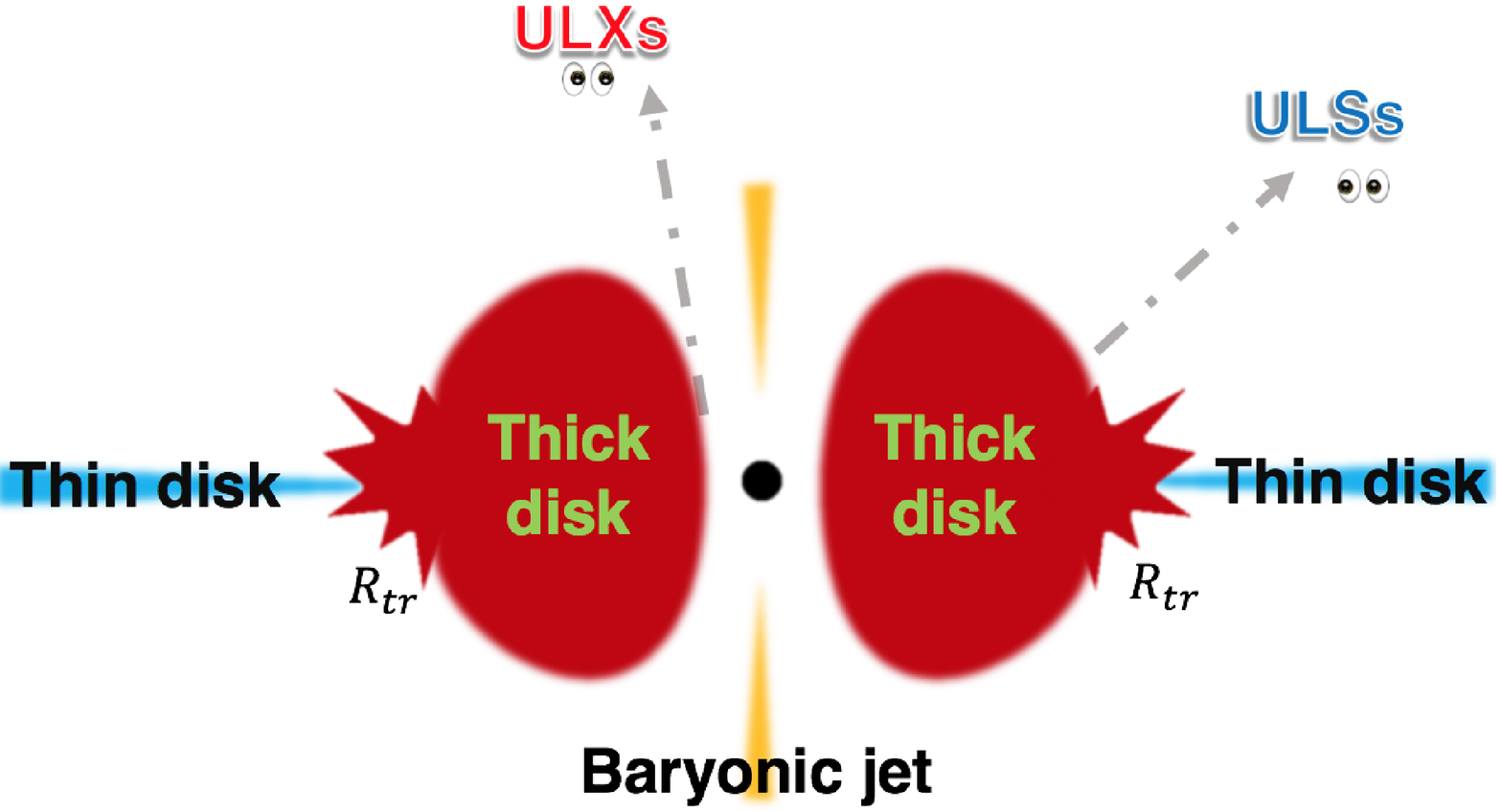}
\caption{
A cartoon picture of super-Eddington accretion around stellar-mass
black holes for $\dot m \ga 30$, which shows ULSs and normal ULXs
with different inclination angles.
}
\end{figure}

\clearpage

\begin{figure}
\plotone{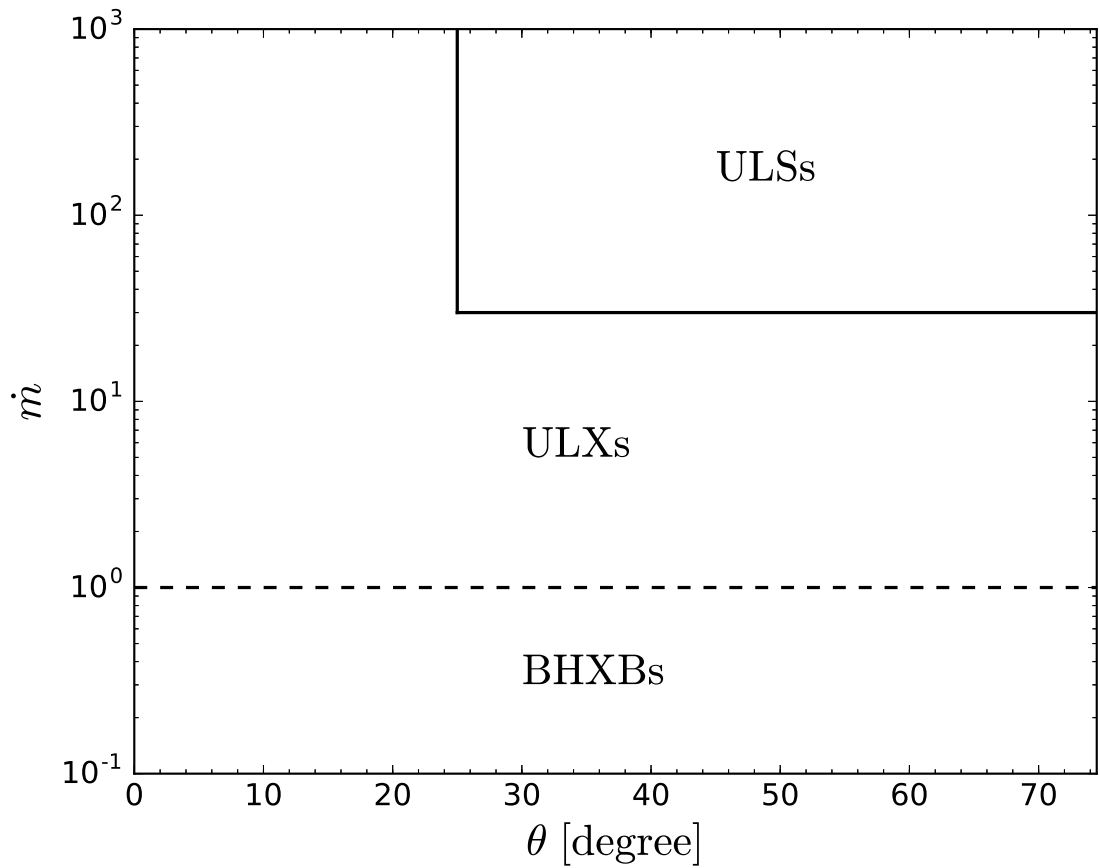}
\caption{
A unified description of ULSs, normal ULXs, and BHXBs based on stellar-mass
black hole systems. BHXBs in our Galaxy are generally below
the Eddington luminosity. ULSs will appear for both high accretion rates
$\dot m \ga 30$ and not small inclination angles $\theta \ga 25^{\circ}$.
For the other cases with super-Eddington accretion, the sources are likely
to be normal ULXs.
}
\end{figure}

\clearpage

\begin{figure}
\plotone{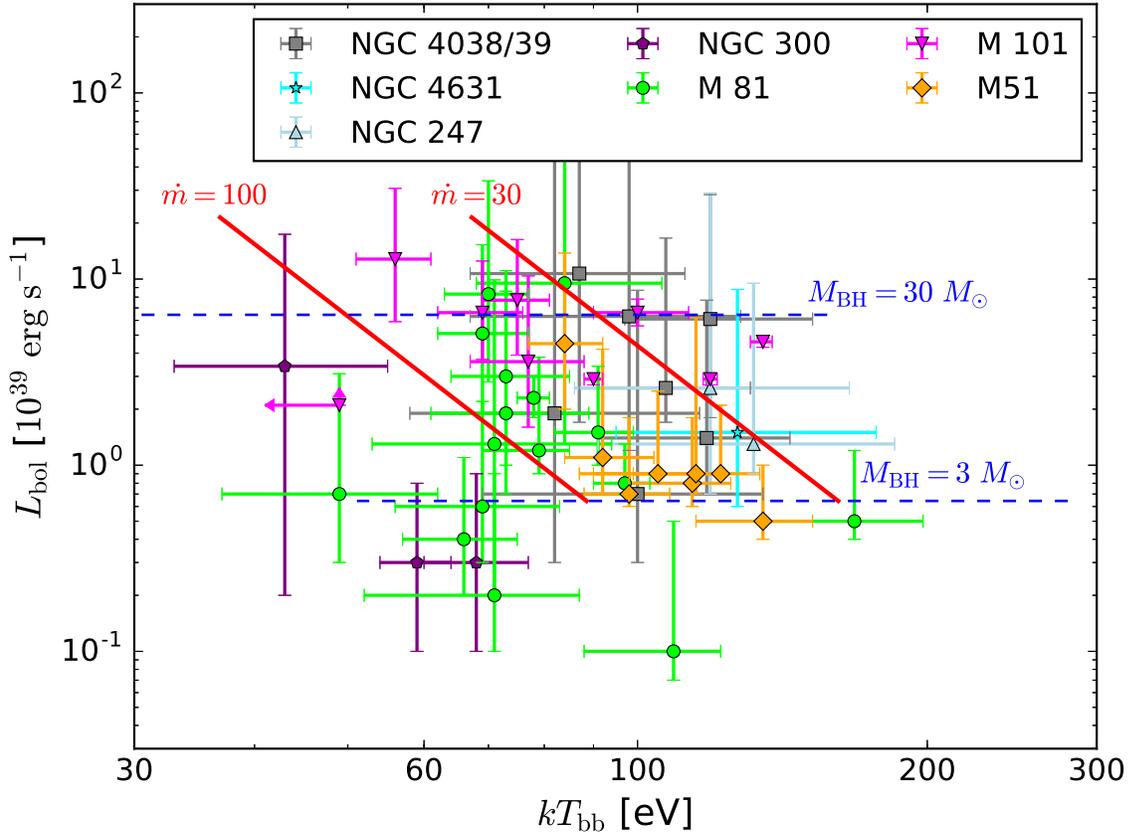}
\caption{
Comparison of the observations (symbols) with our analytic results
(solid and dashed lines) for ULSs in $L_{\rm bol}-kT_{\rm bb}$ diagram.
The symbols represent seven ULSs, which are taken from Table~2
of \citet{Soria15b}.
The horizontal blue dashed lines correspond to fixed black hole masses,
and the inclined red solid lines correspond to fixed mass accretion rates.
}
\end{figure}

\end{document}